\documentclass[preprint2]{aastex}

\slugcomment{astro-ph only, not submitted to any journal}

\shorttitle{The UKIDSS DR2}
\shortauthors{S. J. Warren et al.}

\begin{document}


\title{The UKIRT Infrared Deep Sky Survey Second Data
Release}


\author{S. J. Warren\altaffilmark{1},
N. J. G. Cross\altaffilmark{2},
S. Dye\altaffilmark{3},
N. C. Hambly\altaffilmark{2},
O. Almaini\altaffilmark{4},
A. C. Edge\altaffilmark{5},
P. C. Hewett\altaffilmark{6},
S. T. Hodgkin\altaffilmark{6},
M. J. Irwin\altaffilmark{6},
R. F. Jameson\altaffilmark{7},
A. Lawrence\altaffilmark{2},
P. W. Lucas\altaffilmark{8},
D. J. Mortlock\altaffilmark{1},
A. J. Adamson\altaffilmark{9},
J. Bryant\altaffilmark{2},
R. S. Collins\altaffilmark{2},
C. J. Davis\altaffilmark{9},
J. P. Emerson\altaffilmark{10},
D. W. Evans\altaffilmark{6},
E. A. Gonzales-Solares\altaffilmark{6},
P. Hirst\altaffilmark{9,11},
T. H. Kerr\altaffilmark{9},
J. R. Lewis\altaffilmark{6},
R. G. Mann\altaffilmark{2},
M. G. Rawlings\altaffilmark{9},
M. A. Read\altaffilmark{2},
M. Riello\altaffilmark{6},
E. T. W. Sutorius\altaffilmark{2},
W. P. Varricatt\altaffilmark{9}}
\altaffiltext{1}{Astrophysics Group, Imperial College London, Blackett
  Laboratory, Prince Consort Road, London, SW7 2AZ, U.K.}
\altaffiltext{{2}}{Scottish Universities Physics Alliance (SUPA),
  Institute for Astronomy, School of Physics, University of Edinburgh,
  Royal Observatory, Blackford Hill, Edinburgh, EH9 3HJ, U.K.}
\altaffiltext{3}{Cardiff University, School of Physics \& Astronomy,
  Queens Buildings, The Parade, Cardiff, CF24 3AA, U.K.}
\altaffiltext{4}{School of Physics and Astronomy, University of Nottingham,
  University Park, Nottingham, NG7 2RD, U.K.} 
\altaffiltext{5}{Department of Physics, Durham University, South Road, DH1 3LE,
  U.K.}
\altaffiltext{6}{Institute of Astronomy, Madingley Rd., Cambridge, CB3
  0HA, U.K.} 
\altaffiltext{7}{Department of Physics and Astronomy, University of Leicester,
  Leicester, LE1 7RH, U.K.}
\altaffiltext{8}{Centre for Astrophysics Research, Science and
  Technology Research Institute, University of Hertfordshire,
  Hatfield, AL10 9AB, U.K.} 
\altaffiltext{9}{Joint Astronomy Centre, 660 N. A'ohoku Place, University Park,
  Hilo, Hawaii 96720, U.S.A.}
\altaffiltext{{10}}{Astronomy Unit, School of Mathematical Sciences, Queen Mary,
  University of London, Mile End Road, London E1 4NS, U.K.}
\altaffiltext{{11}}{Gemini Observatory, Northern Operations Center,
  670 North A'ohoku Place, Hilo, HI96720, U.S.A.} 


\begin{abstract}
The UKIRT Infrared Deep Sky Survey (UKIDSS) is a set of five large
near--infrared surveys, covering a complementary range of areas,
depths, and Galactic latitudes. The UKIDSS Second Data Release (DR2)
includes the First Data Release (DR1), with minor improvements, plus
new data for the LAS, GPS, GCS, and DXS, from observations made over
2006 May through July (when the UDS was unobservable). DR2 was staged
in two parts. The first part excluded the GPS, and took place on 2007
March 1. The GPS was released on 2007 April 12.  DR2 includes 282
deg$^2$ of multicolour data to (Vega) $K=18$, complete in the full
{\em YJHK} set for the LAS, 57 deg$^2$ in the {\em ZYJHK} set for the
GCS, and 236 deg$^2$ in the {\em JHK} set for the GPS. DR2 includes
nearly 7 deg$^2$ of deep {\em JK} data (DXS, UDS) to an average depth
$K=21$.  In addition the release includes a comparable quantity of
data where coverage of the filter set for any survey is incomplete. We
document changes that have occurred since DR1 to the pipeline,
calibration, and archive procedures. The two most noteworthy changes
are presentation of the data in a single database (compared to two
previously), and provision of additional error flags for detected
sources, flagging potentially spurious artifacts, corrupted data and
suspected cross-talk sources. We summarise the contents of each of the
surveys in terms of filters, areas, and depths.
\end{abstract}



\section{Introduction}

UKIDSS is the UKIRT Infrared Deep Sky Survey \citep{lawrence07},
carried out using the Wide Field Camera \citep[WFCAM;][]{casali07}
installed on the United Kingdom Infrared Telescope (UKIRT). Data
acquisition for the survey started in 2005 May. The Early Data Release
\citep[EDR;][hereafter D06]{dye06}, a prototype dataset, and the First
Data Release \citep[DR1;][hereafter W07]{war07}, the first release of
survey-quality data, took place in 2006. This paper describes the
Second Data Release (DR2). The data are available from the WFCAM
science archive (WSA) at {\tt http://surveys.roe.ac.uk/wsa}\,.

UKIDSS is a programme of five imaging surveys that each uses some or
all of the broadband filter complement $ZYJHK$, and that span a range
of areas, depths, and Galactic latitudes. There are three high
Galactic latitude surveys, providing complementary combinations of
area and depth; the Large Area Survey (LAS), will cover 4000 deg$^2$
to $K=18$, the Deep ExtraGalactic Survey (DXS), 35 deg$^2$ to
$K=21$, and the Ultra Deep Survey (UDS), 0.8 deg$^2$ to $K=23$. There
are two other wide surveys to $K=18$, aimed at targets in the Milky
Way; the Galactic Plane Survey (GPS) will cover 1900 deg$^2$, and the
Galactic Clusters Survey (GCS) 1100 deg$^2$. The complete UKIDSS
programme is scheduled to take seven years, requiring $\sim$1000
nights on UKIRT. The current implementation strategy is focused on
completing an intermediate set of goals, defined by the `2-year plan'
detailed in D06. All magnitudes quoted in this paper use the Vega
system described by \citet{hewett06}. Depths, where not explicitly
specified, are the total brightness of a point source for which the
flux integrated in a $2\arcsec$ diameter aperture is detected at
$5\sigma$.

A set of five baseline papers provides the relevant technical
background information for the surveys. The overview of the programme
is given by \citet{lawrence07}. This sets out the science goals that
drove the design of the survey programme, and details the final
coverage that will be achieved, in terms of fields, areas, depths, and
filters. The camera is described in detail by \citet{casali07}, and
the {\em ZYJHK} photometric system is characterised by
\citet{hewett06}, who provide synthetic colours for a wide range of
types of star, galaxy, and quasar. Details of the data pipeline and
data archive will appear in Irwin et al. (2007, in prep.) and Hambly
et al. (2007), respectively.

Each UKIDSS data release includes the data contained in previous
releases, together with new data. The instrument is scheduled in
blocks of time of typically three to six months duration. Current
policy is that each observing block results in a data release. DR1
comprised data from the 2005A and 2005B observing blocks. DR2 includes
the relatively short 2006A observing block, which ran over 2006 May to
July. DR3 will be much larger, including the consecutive 2006B and
2007A blocks, running over 2006 November to 2007 May.

Each UKIDSS data release is accompanied by a paper summarising the
contents of the release, and detailing procedural changes since the
previous release. The EDR provided a small prototype datset, and
represented a step towards regular release of survey-quality data.The
EDR paper, D06, is a self-contained summary of all information
relevant to understanding the contents of the EDR. It also serves as
the baseline paper for technical details for all releases. Besides a
summary of the contents of the EDR database, D06 includes relevant
details of the camera design, the observational implementation
(integration times, microstepping), the pipeline and calibration, data
artifacts, and the quality control procedures, as well as a brief
guide to querying the archive. Subsequent papers summarise the
contents of the particular release, together with a description of
changes to the implementation, pipeline, calibration, quality control,
and archive procedures. The DR1 paper, W07, also included plots
summarising the distributions of seeing, sky brightness, and airmass
over the dataset.

The present paper covers DR2. DR2 includes new data for the LAS, GPS,
GCS, and DXS, but not the UDS which was unobservable in 2006A.  The
UDS therefore appears in DR2 in identical form to DR1. Details
concerning the UDS may be found in W07, and no further mention is made
here. DR2 is appeared in two stages. The first stage, released on 2007
March 1, was complete except for the GPS. The GPS data were released
on 2007 April 12. In Section \ref{update} we detail changes between
DR1 and DR2 to the pipeline, calibration, and archive procedures. In
Section \ref{summary} we summarise the contents of DR2.

\section{Update}
\label{update}

D06 contains details of the implementation, pipeline, calibration,
quality control, and archive procedures applied to the EDR data, and a
glossary of technical terms. W07 details changes to these procedures
betwen the EDR and DR1. In this section we detail further changes
since DR1. These include mostly minor changes to the pipeline,
calibration, and archive procedures. The implementation and quality
control procedures are virtually unchanged. During 2006A, fields were
not observed when the moon was within 30deg, in order to avoid moon
ghosts (D06, W07). A baffle was installed after the end of 2006A that
appears to have eliminated moon ghosts altogether, and the restriction
has been removed for 2006B.

\subsection{Pipeline and calibration}

\subsubsection{Pipeline}
\label{pipe}

For DR2 all 2005A WFCAM data were reprocessed from scratch from the
raw data. The 2005A data were first released in the EDR, and were not
revised for DR1. The main changes in the image processing aspects
between EDR and DR1, and used in processing the 2005B data, were an
improved sky correction strategy more closely linked to the observing
block structure (MSBs), and the application of the cross-talk
supression algorithm (W07). Further minor improvements have been made
to the sky correction strategy since DR1. Therefore the 2005A and
2006A data benefit from these improvements. The 2005B data have not
been reprocessed since DR1 however.

Further small changes to the cataloguing software, mainly related to
improving the object detection filter, have also been made since DR1 and 
were used in the catalogue (re)generation for the 2005A data. All of the new 
2006A WFCAM data were processed with the latest versions of the pipeline 
software.

Several minor improvements to the photometric calibration procedure have
also been implemented since DR1 in an attempt to reduce any dependence of
the derived photometric zero-points on Galactic coordinates, due to, for 
example, varying dwarf-giant ratios, extinction and extreme colour objects.

As previously, each frame is calibrated using 2MASS stars in the
frame, converting the 2MASS photometry to the WFCAM system using
appropriate colour equations. The zero point for each detector is now
provided as the attribute {\tt photZPCat} in the table {\tt
  MultiframeDetector}. The attribute {\tt photZP} in the table {\tt
  Multiframe} has been deprecated.  The updated WFCAM calibration uses
a restricted (extinction-corrected) colour range of $0 < J_2-K_2 < 1$
(the subscript $2$ standing for 2MASS) to help exclude late-type
giants, unusual objects and heavily reddened stars. Also, as a result
of discovering a correlation of the derived $Y$- and $Z$-band zero
points with Galactic extinction, particularly in regions of heavy
extinction, an extra extinction-dependent calibration term has been
included. The revised colour equations are as follows: \\
$Z=J_2+ 0.95(J_2-H_2)+0.39E(B-V)'$ \\ 
$Y=J_2+ 0.50(J_2-H_2)+0.16E(B-V)'$ \\
$J=J_2-0.065(J_2-H_2)+0.015E(B-V)'$ \\
$H=H_2+0.07(J_2-H_2)+0.005E(B-V)'-0.03$ \\
$K=K_2+0.01(J_2-K_2)+0.005E(B-V)'$  

Here $E(B-V)'$ is the reddening computed using the prescription of
\citet{bonif}, from the data of \citet{schlegel}. The corrections for
reddening are small in the {\em JHK} bands and in practice also for the
$Y$ band which is only used in the LAS where the extinction is low.  The
reddening corrections for the {\em Z} band, only used in the GCS, are
large in fields with large reddening, and should be treated with
caution.

The revised colour equations produce a more stable calibration for
the $Y$ and $Z$ bands (but generally have little impact on the {\em JHK}
calibration).  In addition, errors on the individual 2MASS-derived
zero-points are now included together with measures summarising the
photometric quality of the night and the overall nightly average
zero-point. 

W07 note that there is evidence for an overall zero-point offset with
respect to the Vega system for the $Y$ and $Z$ bands.
An analysis of this issue is deferred to DR3.

\begin{figure*}
\includegraphics[width=17cm]{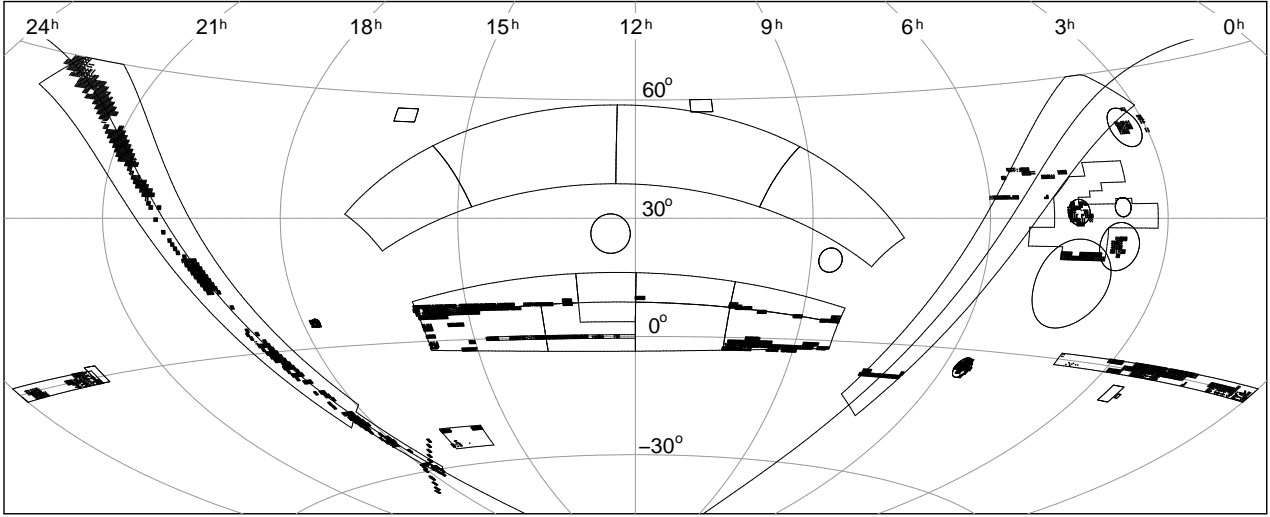}
\caption{Plot showing regions with full filter coverage for the LAS
  ({\em YJHK}), GCS ({\em ZYJHK}), and GPS ({\em JHK}). The outlines
  are the boundaries of the surveys (see \citet{lawrence07}, and D06
  for details), and the filled regions are the fields included in
  DR2. The three LAS zones are the thin equatorial stripe, and the two
  broad bands in the centre. The GPS fields are confined to the large
  parabola, centred on the Plane, plus the Taurus-Auriga-Perseus
  region, with the complicated outline, near 4${\rm ^h}$30${\rm ^m}$, +30deg. All other
  regions containing data are GCS fields.}
\label{coverage_dr2}
\end{figure*}

\begin{figure*}
\includegraphics[width=17cm]{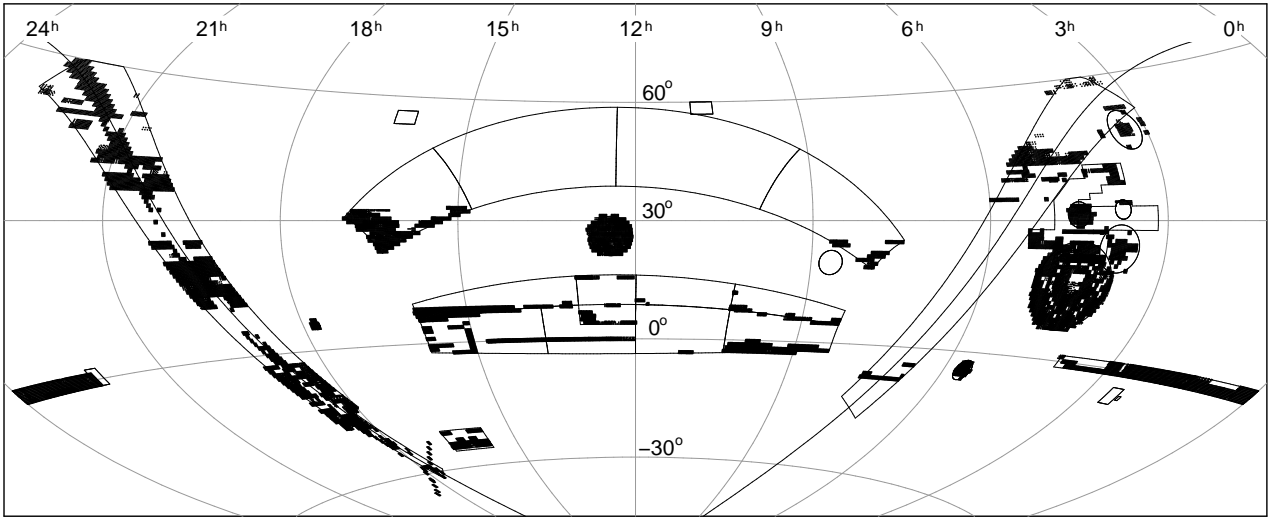}
\caption{Plot showing regions of the LAS and GCS with coverage with
  any filters. The outlines are the boundaries of the surveys (see
  \citet{lawrence07}, and D06 for details), and the filled regions are
  the fields included in DR2. The three LAS regions are the thin
  equatorial stripe, and the two broad bands in the centre.
  Observations in the northern LAS block are first-epoch {\em J}-band
  only. The GPS fields are confined to the large parabola, centred on
  the Plane, plus the Taurus-Auriga-Perseus region, with the
  complicated outline, near 4${\rm ^h}$30${\rm ^m}$, +30deg. All other regions
  containing data are GCS fields.}
\label{coverage_dr2+}
\end{figure*}

\subsection{Archive}
\label{archivesec}

In previous releases two databases were provided, one containing all
the data (called EDR+, DR1+), and a second, a subset of the larger database,
containing data for all fields observed with the full filter
complement for the particular survey (called EDR, DR1). For DR2 there is only
one database, DR2+. Nevertheless {\em views} are provided of the
source tables for each survey, e.g. {\em lasYJHKsource} in order to
provide the equivalent functionality.

A significant improvement over DR1 is the provision of quality error
bit flags for each detected source. In the detection tables the
attribute is {\tt ppErrBits}, and in the source tables the appropriate
filter name is appended, e.g.~{\tt yppErrBits, j$\_$1ppErrBits}. The
attribute is a 32-bit integer where each bit can be set to 1 to flag
up to 32 different quality issues. The quality issues are ranked in
severity, with the more severe issues assigned to higher
bits. Currently only five bits are set as follows: Bit 4 = the source
is the offspring of deblending, Bit 6 = a bad pixel exists in the
default aperture ($1\arcsec$ radius), Bit 16 = the source is likely to
contain saturated pixels, Bit 19 = the source is likely to be either
cross-talk or a real source affected by cross-talk (see D06 and W07
for details on cross-talk), Bit 22 = the source is likely to include
incomplete data as a consequence of lying close to the frame
boundary. Therefore by converting {\tt ppErrBits} to binary it is
possible to filter out sources with particular quality
issues. Alternatively a threshold can be set e.g. {\tt
  ppErrBits$<32$} to select sources with only very minor quality
issues.\footnote{Further details are provided at \\ {\tt
  http://surveys.roe.ac.uk/wsa/drtwo/ppErrBits.html}}

In addition the seaming algorithm has been improved since DR1. The
seaming algorithm deals with sources that are detected in separate
(two or more) multiframes that overlap. To decide which detection will
be designated primary a sequence of discriminants is considered, until
a preference is established: first the number of filters in which the
source is detected, second the count from {\tt ppErrBits}, and finally
the distance of the source from the detector edge.

UKIDSS DR2 is crossmatched with SDSS DR5 \citep{sdssdr5}, updated from
SDSS DR3 for UKIDSS DR1. Note also that new crossmatching with
catalogues from the Millenium Galaxy Catalogue \citep{mgc} and the
NRAO VLA Sky Survey \citep{nvss} is provided for DR2.

\begin{deluxetable}{ccccccccc}
\tablecaption{Coverage of the shallow surveys (deg$^2$) in DR2. For the GPS
  `all filters' implies all of {\em JHK}.\label{tab_shallow_coverage}}
\tablehead{
Survey & all     & \multicolumn{7}{c}{filter} \\
       & filters &  $Z$ &  $Y$ &  $J$ &  $H$ &  $K$ & H$_2$ & any }
\tablewidth{0pt}
\startdata
LAS    & 282 &  -  & 341 & 505 & 477 &  476 &   - &  685 \\
GCS    &  57 &  80 &  81 &  83 & 116 &  478 &   - &  496 \\
GPS    & 236 &  -  &  -  & 249 & 252 &  657 &  70 &  676 \\ \hline
total  & 575 &  80 & 422 & 837 & 845 & 1611 &  70 & 1857 \\ \hline
\enddata
\end{deluxetable}

\begin{deluxetable}{cccc}
\tablecaption{The median $5\sigma$ point source depth by filter in the DR2
  database for the shallow surveys.\label{tab_shallow_depth}}
\tablewidth{0pt}
\tablehead{
Filter & LAS & GCS & GPS}
\startdata
$Z$ &   -    & 20.54 &  -  \\
$Y$ & 20.23  & 20.16 &  -  \\
$J$ & 19.61  & 19.59 & 19.77 \\
$H$ & 18.90  & 18.87 & 19.00 \\
$K$ & 18.23  & 18.20 & 18.05 \\
\enddata
\end{deluxetable}

\section{Summary of the contents of DR2}
\label{summary}

\subsection{LAS, GCS, GPS}

For the shallow surveys, LAS, GCS, GPS, Figs \ref{coverage_dr2} and
\ref{coverage_dr2+} plot the survey coverage in
DR2. Fig. \ref{coverage_dr2} illustrates the sky coverage of fields
where the filter set for the particular survey is complete, while
Fig. \ref{coverage_dr2+} shows all fields in which any data have been
taken. The new LAS data are confined to the region $10^{\mbox
  {\scriptsize h}}53^{\mbox {\scriptsize m}}<RA<16^{\mbox {\scriptsize
    h}}50^{\mbox {\scriptsize m}}$ The new observations in the northern
block, near $RA=16^{\mbox {\scriptsize h}}$, are J-band only, and are
first epoch observations for proper motions in fields that will later
be reobserved in the full filter complement, after a minimum interval
of two years (D06).

Details of the summed area covered, by filter, in each of the shallow
surveys, LAS, GCS, and GPS, are provided in Table
\ref{tab_shallow_coverage}. For each survey, the table provides the
area covered in a particular filter, the area covered by all filters,
and the area covered by any filter. Table \ref{tab_shallow_depth}
provides the median $5\sigma$ depth (as defined in D06) achieved in
each band in each of the two surveys. These values are slightly deeper
than those for DR1, provided in W07.

A quirk persists from DR1 (see section 4.1 in W07) which results in a
small number of LAS sources recorded in the archive as detected in {\em
  YJ} only, coincident with distinct sources (in reality the same
sources) recorded as detected in {\em HK} only.

\begin{deluxetable}{lccccccccccc}
\tablecaption{The DXS {\em deepleavstack} multiframes. The field, sub-field
  code, and base coordinates are listed, then, for $J$ and $K$
  successively, the total integration time, 
  5$\sigma$ depth, seeing, and ellipticity
  $e=1-b/a$.\label{tab_dxs_deep_stacks}}
\tabletypesize{\footnotesize}
\tablewidth{0pt}
\tablehead{
Field &  Sub & RA & Dec & $t_{tot}$ & Depth & seeing & ellip. &
$t_{tot}$ & Depth & seeing & ellip. \\ 
 & field & \multicolumn{2}{c}{J2000} & s & mag. & $\arcsec$ & & s & mag. &
$\arcsec$ & \\
 & & & & \multicolumn{4}{c}{{\em J}} & \multicolumn{4}{c}{{\em K}}
}
\startdata 
XMM-LSS      & 1.00 &  36.5752020 & -4.7496444 &  6400 & 22.195 & 0.88 & 0.05 &  7040 & 20.845 & 0.76 & 0.08 \\
XMM-LSS      & 1.10 &  36.5803542 & -4.5294528 &  7680 & 22.265 & 0.86 & 0.06 &  8320 & 20.935 & 0.74 & 0.09 \\
XMM-LSS      & 1.01 &  36.8012000 & -4.7496444 &  7680 & 22.245 & 0.86 & 0.05 &  7680 & 20.885 & 0.77 & 0.08 \\
XMM-LSS      & 1.11 &  36.8012000 & -4.5294528 &  7040 & 22.275 & 0.86 & 0.05 &  5760 & 20.735 & 0.75 & 0.09 \\
XMM-LSS      & 2.00 &  35.7098417 & -4.7496444 &   -   &   -    &  -   &  -   &  6400 & 20.675 & 0.93 & 0.08 \\
XMM-LSS      & 2.10 &  35.7098417 & -4.5294528 &   -   &   -    &  -   &  -   &  7040 & 20.715 & 0.91 & 0.07 \\
XMM-LSS      & 2.01 &  35.9306875 & -4.7496444 &   -   &   -    &  -   &  -   &  7040 & 20.695 & 0.91 & 0.08 \\
XMM-LSS      & 2.11 &  35.9306875 & -4.5294528 &   -   &   -    &  -   &  -   &  6400 & 20.675 & 0.89 & 0.07 \\
XMM-LSS      & 3.00 &  35.7094541 & -3.8688833 &   -   &   -    &  -   &  -   &   640 & 19.325 & 1.17 & 0.11 \\
XMM-LSS      & 3.01 &  35.9300625 & -3.8688833 &   -   &   -    &  -   &  -   &   640 & 19.375 & 1.12 & 0.10 \\ \hline
Lockman Hole & 1.00 & 163.3623000 & 57.4753556 &  1280 & 21.565 & 1.06 & 0.10 & 12600 & 20.995 & 0.93 & 0.09 \\
Lockman Hole & 1.10 & 163.3623000 & 57.6955472 &  1280 & 21.575 & 1.06 & 0.08 & 13960 & 21.055 & 0.98 & 0.09 \\
Lockman Hole & 1.01 & 163.7756170 & 57.4753556 &   640 & 21.085 & 1.16 & 0.06 & 13820 & 21.025 & 0.99 & 0.08 \\
Lockman Hole & 1.11 & 163.7756170 & 57.6955472 &  1280 & 21.465 & 1.15 & 0.08 & 11820 & 20.965 & 0.98 & 0.08 \\
Lockman Hole & 6.00 & 161.6707620 & 59.1223000 &   -   &   -    &  -   &  -   &  8840 & 20.765 & 1.06 & 0.08 \\
Lockman Hole & 6.10 & 161.6707620 & 59.3424917 &   -   &   -    &  -   &  -   &  9840 & 20.855 & 1.05 & 0.08 \\
Lockman Hole & 6.01 & 162.1040380 & 59.1223000 &   -   &   -    &  -   &  -   &  9340 & 20.815 & 1.05 & 0.06 \\
Lockman Hole & 6.11 & 162.1040380 & 59.3424917 &   -   &   -    &  -   &  -   &  8700 & 20.745 & 1.06 & 0.08 \\ \hline
ELAIS N1     & 1.00 & 242.5994000 & 54.5031333 &  6560 & 21.785 & 1.01 & 0.06 &  8000 & 20.675 & 0.99 & 0.08 \\
ELAIS N1     & 1.10 & 242.5994000 & 54.7233250 &  6060 & 21.785 & 0.96 & 0.07 &  8500 & 20.795 & 0.98 & 0.08 \\
ELAIS N1     & 1.01 & 242.9817370 & 54.5031333 &  8480 & 21.855 & 0.99 & 0.07 &  9000 & 20.845 & 0.98 & 0.09 \\
ELAIS N1     & 1.11 & 242.9817370 & 54.7233250 &  9480 & 21.845 & 1.04 & 0.07 &  8000 & 20.725 & 0.97 & 0.09 \\
ELAIS N1     & 2.00 & 241.1580080 & 54.5031333 &  8320 & 22.125 & 0.92 & 0.05 &  9400 & 20.815 & 0.89 & 0.08 \\
ELAIS N1     & 2.10 & 241.1580080 & 54.7233250 &  7680 & 22.075 & 0.93 & 0.05 &  7620 & 20.595 & 0.91 & 0.09 \\
ELAIS N1     & 2.01 & 241.5403460 & 54.5031333 &  8960 & 22.115 & 0.94 & 0.05 &  9260 & 20.785 & 0.89 & 0.08 \\
ELAIS N1     & 2.11 & 241.5403460 & 54.7233250 &  6400 & 21.955 & 0.94 & 0.05 &  8400 & 20.715 & 0.87 & 0.08 \\
ELAIS N1     & 3.00 & 241.1364170 & 55.3170222 &   -   &   -    &  -   &  -   &   500 & 19.165 & 1.08 & 0.05 \\
ELAIS N1     & 3.10 & 241.1364170 & 55.5372139 &   -   &   -    &  -   &  -   &   500 & 19.215 & 1.09 & 0.05 \\
ELAIS N1     & 3.01 & 241.5266580 & 55.3170222 &   -   &   -    &  -   &  -   &   500 & 19.275 & 0.96 & 0.09 \\
ELAIS N1     & 3.11 & 241.5266580 & 55.5372139 &   -   &   -    &  -   &  -   &   500 & 19.215 & 0.95 & 0.12 \\
ELAIS N1     & 4.00 & 242.6111380 & 55.3170222 &   -   &   -    &  -   &  -   &   500 & 19.405 & 0.83 & 0.09 \\
ELAIS N1     & 4.10 & 242.6111380 & 55.5372139 &   -   &   -    &  -   &  -   &   500 & 19.415 & 0.81 & 0.11 \\
ELAIS N1     & 4.01 & 243.0013830 & 55.3170222 &   -   &   -    &  -
&  -   &   500 & 19.405 & 0.86 & 0.09 \\
\tablebreak
VIMOS 4      & 1.00 & 334.2667460 &  0.1698000 &  8820 & 22.175 & 0.89 & 0.04 & 10400 & 20.945 & 0.83 & 0.06 \\
VIMOS 4      & 1.10 & 334.2667460 &  0.3899917 & 10740 & 22.285 & 0.86 & 0.04 & 12820 & 21.085 & 0.79 & 0.06 \\
VIMOS 4      & 1.01 & 334.4869460 &  0.1698000 & 10100 & 22.355 & 0.84 & 0.04 & 10900 & 21.055 & 0.80 & 0.06 \\
VIMOS 4      & 1.11 & 334.4869460 &  0.3899917 &  6900 & 22.165 & 0.88 & 0.05 & 10260 & 20.975 & 0.85 & 0.06 \\
VIMOS 4      & 2.00 & 335.1420250 &  0.1817444 &  5120 & 21.855 & 0.92 & 0.04 &  8960 & 20.945 & 0.75 & 0.06 \\
VIMOS 4      & 2.10 & 335.1420250 &  0.4019361 &  4480 & 21.845 & 0.93 & 0.03 & 10240 & 20.965 & 0.74 & 0.06 \\
VIMOS 4      & 2.01 & 335.3622250 &  0.1817444 &  5120 & 21.845 & 1.02 & 0.04 & 10240 & 20.945 & 0.74 & 0.06 \\
VIMOS 4      & 2.11 & 335.3622250 &  0.4019361 &  5760 & 21.925 & 0.94 & 0.04 &  9600 & 20.885 & 0.75 & 0.06 \\
VIMOS 4      & 3.00 & 335.1420790 &  1.0559111 & 11520 & 22.325 & 0.85 & 0.04 & 10240 & 20.795 & 0.94 & 0.06 \\
VIMOS 4      & 3.10 & 335.1420790 &  1.2761028 & 11520 & 22.375 & 0.87 & 0.04 & 11520 & 20.905 & 0.98 & 0.05 \\
VIMOS 4      & 3.01 & 335.3623330 &  1.0559111 &  9600 & 22.315 & 0.88 & 0.04 & 11520 & 20.885 & 0.98 & 0.05 \\
VIMOS 4      & 3.11 & 335.3623330 &  1.2761028 & 11520 & 22.365 & 0.88 & 0.05 & 10880 & 20.895 & 0.93 & 0.06 \\
VIMOS 4      & 4.00 & 334.2668040 &  1.0559111 &  1920 & 21.535 & 0.98 & 0.04 & 10240 & 20.915 & 0.85 & 0.06 \\
VIMOS 4      & 4.10 & 334.2668040 &  1.2761028 &  2560 & 21.725 & 0.93 & 0.03 & 12160 & 20.975 & 0.87 & 0.05 \\
VIMOS 4      & 4.01 & 334.4870580 &  1.0559111 &  1920 & 21.515 & 1.00 & 0.04 & 14720 & 21.085 & 0.86 & 0.06 \\
VIMOS 4      & 4.11 & 334.4870580 &  1.2761028 &  1280 & 21.285 & 0.93 & 0.04 & 12800 & 21.035 & 0.87 & 0.06 \\
VIMOS 4      & 5.00 & 333.3940250 &  1.0559111 &   -   &   -    &  -   &  -   &   640 & 19.345 & 1.03 & 0.12 \\
VIMOS 4      & 5.10 & 333.3940250 &  1.2761028 &   -   &   -    &  -   &  -   &   640 & 19.235 & 1.15 & 0.13 \\
VIMOS 4      & 6.00 & 333.3939708 &  0.2064666 &   -   &   -    &  -   &  -   &   640 & 19.285 & 1.15 & 0.10 \\
VIMOS 4      & 6.10 & 333.3939708 &  0.4266583 &   -   &   -    &  -   &  -   &   640 & 19.385 & 1.06 & 0.05 \\
VIMOS 4      & 6.01 & 333.6141708 &  0.2064666 &   -   &   -    &  -   &  -   &   640 & 19.215 & 1.10 & 0.03 \\
VIMOS 4      & 6.11 & 333.6141708 &  0.4266583 &   -   &   -    &  -   &  -   &   640 & 19.365 & 1.05 & 0.03
\enddata
\end{deluxetable}

\subsection{DXS}

The DXS is targeting the following four fields: XMM-LSS (2$^{\rm
  h}25^{\rm m}$,$-4^{\circ}30'$), the Lockman Hole (10$^{\rm h}57^{\rm
  m}$,+57$^{\circ}40'$), Elais N1 (16$^{\rm h}10^{\rm
  m}$,+54$^{\circ}00'$), and VIMOS 4 (22$^{\rm h}17^{\rm
  m}$,+0$^{\circ}20'$). The fields are plotted as small quadrilaterals
in Figs \ref{coverage_dr2} and \ref{coverage_dr2+}. Eventually each
field will be tessellated by 12 tiles, each of four pointings. The
observing strategy \citep{lawrence07} aims to reach full depth in a
tile, $K=21$, $J=22.5$, before moving to the next tile. Depth is built
up by a series of observations of length 640s (500s in 2005A), known
as intermediate stacks. All the intermediate stacks for any pointing
are combined to create {\em deepleavstack} multiframes (D06). All
intermediate stack and {\em deepleavstack} multiframes are included in
a release.

In DR2 the numbers of {\em deepleavstack} multiframes are 55 in $K$
and 32 in $J$. Details of all the DXS {\em deepleavstacks} are
provided in Table \ref{tab_dxs_deep_stacks}. The first column provides
the field name, and the second is a code identifying the sub-field, in
the form {\em Tile.XY}, where {\em XY} are binary coordinates
specifying one of four multiframe positions that make up a
tile. Columns three and four provide the coordinates of the base
position, which is the centre of the field of view (which is a point
not imaged by the detectors).  The remaining columns give the total
integration time $t_{\rm tot}$ of the frames contributing to the stack
(i.e. excluding deprecated frames), the $5\sigma$ depth, the seeing,
and the ellipticity, for the {\em J} and {\em K} frames.

In total the DXS contains fields with deep {\em JK} coverage, reaching
$K\sim21$, over 6\,deg$^2$. 

\subsection{Known issues with DR2}

The Release History web page \\
{\tt http://surveys.roe.ac.uk/wsa/releasehistory.html} provides an
up-to-date summary of known problems with DR2.

\setlength{\bibhang}{2.0em}

\end{document}